# Design and Development of a Nanoscale Multi Probe System Using Open Source SPM Controller and GXSM Software: A Tool of Nanotechnology.

S. K. Suresh Babu [a], J. S. DevrenjithSingh [c], D. Jackuline Moni [b], D. Devaprakasam [a*]

[a] Faculty, NEMS/MEMS/NANOLITHOGRAPHY Lab, Department of Nanosciences and Technology, Karunya University, Coimbatore-641114, India

[b] Faculty, Department of Electronics and Communication Engineering, Karunya University, Coimbatore-641114, India

[c] PG Scholar, NEMS/MEMS/NANOLITHOGRAPHY Lab, Department of Nanosciences and Technology, Karunya University, Coimbatore-641114, India

**Abstract:**
We report our design, development, installation and troubleshooting of an open source Gnome X Scanning Microscopy (GXSM) software package for controlling and processing of modern Scanning Probe Microscopy (SPM) system as a development tool of Nanotechnology. GXSM is a full featured analysis tool for the characterization of nanomaterials with different controlling tools like Atomic Force Microscopy (AFM), Scanning Tunneling Spectroscopy (STS), scanning tunneling microscopy (STM), Nanoindentation and etc.,. This developed package tool consists of Digital Signal Processing (DSP) and image processing system of SPM. A digital signal processor (DSP) subsystem runs the feedback loop, generates the scanning signals and acquires the data during SPM measurements. With installed SR-Hwl plug-in this developed package was tested in no hardware mode.

## 1. Introduction

Gnome X Scanning Microscopy (GXSM) is a best and powerful tool for data acquisition and controlling of scanning probe microscopy. This tool is used with scanning tunneling microscopy (STM), atomic force microscopy (AFM), Scanning Tunneling Spectroscopy (STS). GXSM provides various methods for 2D data (of various types: byte, short, long, double) visualization and manipulation. We are currently using it for scanning tunneling microscopy (STM), atomic force microscopy (AFM) [1] [2] [3]. Data presentation is by default a (grey or false color) image but it can be switched to a profile view (1d), profile extraction on the fly... Or you can use a 3D shaded view (using MesaGL) which now offers a sophisticated scene setup. The "high-level" scan controller is now separated from the GXSM core and is built as Plug-in, while the real-time "low-level" scanning process, data-acquisition and feedback loop (if needed), runs on the DSP if present, else a dummy image is produced. [4]

Extremely flexible configuration of user settings and data acquisition and probe modes. Special instrument control Plug-Ins. A Plug-in categorizing mechanism automatically only load the required Plug-Ins for the actual setup: E.g. no Hardware Control Plug-ins is loaded in"offline" Data Analysis Mode. There are more than 80 Plug-ins used. GXSM itself is fully hardware independent. It provides a generic hardware-interface (HwI) plugin infrastructure to attach any kind of hardware. The HwI has to manage low level tasks and pass the data to the GXSM core, plus, it has to provide the necessary GUI to provide user access and control to hardware/instrument specific parameters and tasks.

The GXSM software can be divided into three parts: First, the GXSM core providing the main functionality for handling and visualization of data. The basic functions of the GXSM core can be extended using plug-ins. Plug-ins are small pieces of software dynamically linked to the core.

*Tel: 0422-2641300, Fax: 0422-2615615, Email: *devaprakasam@karunya.edu*

The plug-ins is described in the second part of the manual. The third part documents the digital signal processing

(DSP) software needed to carry out actual measurements. The DSP software is not necessary for applications using GXSM only for data analysis purposes.

## 2. Hardware used

*2.1 Signal Ranger MK2-A810*

The Signal Ranger Mark2 DSP with the A810 analog board is powered by the TMS320C5502 running at 300 MHz and uses a high speed USB-2.0 interface for all communications with the host PC and GXSM. The SPM Open Source Controller model used is Mk2-A810. This DSP-based system has been specially designed to meet the Scanning Probe Microscopy (SPM) application requirements. [6]

Advanced SPM features can be implemented using the 16 individually configurable GPIOs and the two16-bit counters. These counters are synchronized with the analog sampling and can be used as simple pulse counters or Quadrature Encoder Pulse (QEP) counters. Based on the SR-MK2 DSP board and SR2-A810 board, the SPM Open Source Controller is a convenient rack-mount enclosure providing quality connectors and wiring to ensure the best S/N ratio. With all these features the kit SR2-A810 + SR-Mk2 has the best performance/price ratio on the market for an SPM control system.

1. USB

    USB 2.0 PC connection. Average data throughput: 18Mb/s (reads), 22Mb/s (writes). The stand-alone USB controller requires no management from the DSP software.

2. DSP

    TMS320C5502 16-bits fixed point DSP, running at300 MHz, with 32Kwords of on-chip RAM.

3. FPGA

    XC3S400 FPGA. 400 000 gates. 56 Kbits distributed RAM, 288 Kbits block RAM, 16 dedicated 18x18 multipliers, 4 DCMs. Provides63 user-configurable I/Os.
4. Power supply

    Signal Ranger_mk2 is Self-Powered using an external 5V (+-5%) power pack. It can work without any connection to a PC.

5. Memory

    - 64 Kbytes on-chip (DSP) double-access RAM, mapped in data and program spaces.
    - 4 Mbytes external 75MHz Synchronous Dynamic RAM, mapped into data and program space.
    - 2 Mbytes external Flash Rom, mapped into data and program space.

*2.2 Piezo Control Systems Design*

The piezo amplifier, NV 40/3 CLE, was designed for low voltage piezo elements with external measurement systems. The NV 40/3 CLE includes a voltage amplifier, closed loop control, and a display with a PC-interface. The high-resolution display shows the position of the actuator, along with some other information such as: the mode of operation and the temperature inside the amplifier. This piezo amplifier also provides the opportunity to operate the piezo element via an analog modulation input. The position of the actuator can be examined via the monitor output. Due to the very low voltage noise of the output current, only 0.3mVRMS, this amplifier system is ideally suited for positioning applications with sub-nm resolution. Special protective circuits prevent voltage spikes when switching the unit on and off and consequently averts any overload caused by overheating or short-circuiting. The new soft start ensures an actuator-safe activation of the system. With the electronic PID controller this system operates without any drift or hysteresis. Due to its design, and compact casing, the new amplifier is recommended for all

R&D and engineering applications.

## 3. GXSM Project Installation

*3.1 System requirements*

GXSM needs a reasonable up-to-date machine running a recent version of almost any Linux Variant. Recommended the Debian based Ubuntu. System memory requirements are ranging from little to several Gigabytes. To compile and run GXSM the full Gtk+/Gnome libraries, development packages and some additional libraries (FFTW, NetCDF, libQuicktime) are required, but all are available as pack- ages for Debian and Ubuntu. GXSM built on a Debian, Ubuntu, Fedora, RedHat, Suse, Mandrake Linux. The re- cent Ubuntu is the GXSM reference system and is used for development [5]. For more Linux- Distribution dependent installation guides and hints, please refer to the online [7] [8]

"GXSM Installation Knowledge" forum:
http://sourceforge.net/p/gxsm/discussion/297458.

*3.1.1 Compiling and installing GXSM*

Change into the Gxsm-2.0 directory in the terminal (ctrl+alt+t). The following steps for compiling Gxsm 2.0 and Sranger.

Configure, Compile, and Install Gsxm-2.0

```
$ cd
$ cd Gxsm-2.0
$ ./autogen.sh
$ make
$ sudo make install
```

Configure, Compile, and Install Sranger

```
$ cd
$ cd SRanger
$ ./autogen.sh
$ make -i
$ sudo make -i install
```

Then install the kernel module for MK2-A810

```
$ cd modules-mk2-2.6.x$ make -C /usr/src/linux-headers-  2.6.35-25-generic/ SUBDIRS=$PWD modules
$ sudo checkinstall make -C /usr/src/linux-headers-2.6.35-25-generic/
SUBDIRS=$PWD modules_install
$ sudo depmod -a
$ sudo modprobe usb-sranger_mk2
```

Then I open "/lib/udev/rules.d/50-udev-default.rules" and added the following comment at the end of file:
KERNEL=="sranger_mk2_" MODE="0666".

Programming DSP and FPGA

From a Windows machine, start SR3_MiniDebugger.exe.  Use SR2_NG in the pull-down menu for the SR2 "NewGeneration"

The two files are required:
1. DSP File: FB_spmcontrol.out (found at /home/fimadmin/SRanger/TiCC-project-files/MK2A810_FB_spmcontrol)
2. FPGA File: SR2_Analog_810_V200.rbt (found at /home/fimadmin/SRanger/TiCC-project-files/MK2A810_FB_spmcontrol/SR2_PrepareFlashForSR2_Analog_810)
The SR firmware tools (includes SR3_MiniDebugger.exe) can be found at
http://www.softdb.com/dsp-products-SR-MK2.php
see the section on SR2_NG
http://www.softdb.com/SignalRangerMk2/SR3_Applications_Installer_V133.exe

If GXSM does not work

To activate the hardware
$ cd ~/SRanger/TiCC-project-files/MK2-A810_FB_spmcontrol/python_scripts
$ ./mk2_spm_control.py
Then
$ gxsm2 --force-configure
Specify the correct hardware type, "SRangerMK2:SPM" and the device file "/dev/sranger_mk2_0".

*3.1.2 Sranger connections*

Make sure the MK2-A810's flash is up-to date with the latest DSP code for Gxsm and (FPGA) logic matching the Gxsm version. (Best to do CVS checkout about the same time for the CVS modules Gxsm-2.0 (Gxsm project) and SRanger modules from Srangerproject.)

**4. Results and Discussion**

4.1 Main Window

The main window provides two different functions: Firstly, it has a menu bar with pull down menus. These menus provide the user with the usual File and Help menus, which can be found in practically every mouse-driven software piece. Some of these pull-down menus interact with (Math) or start-up (Windows) other windows. Secondly, the main menu contains a large number of control fields which can be used, e.g., to control an instrument, or just display certain parameters. These control fields are described in the following two sections. In the main window we can make the changes to the controller option by using the preference tab. All the main changes are being made in this tab.

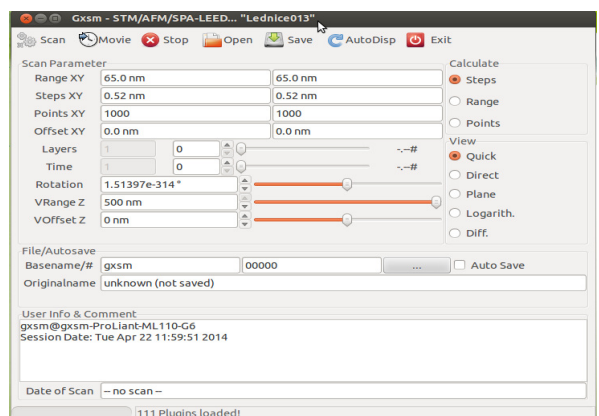

Figure 3 The Main window

### 4.2 SPM Connection Preview Mode:

The SPM connection preview mode is given that SPM controller is connected and the right corner of the window shows the offset of X, Y, Z in the Piezo drive. The channel selector will allow to select the required topography to be scanned whether 1D, 2D or3D need to be scanned. The scan button will start the can and data's are saved as .nc (NetCDF) extension in the home directory.

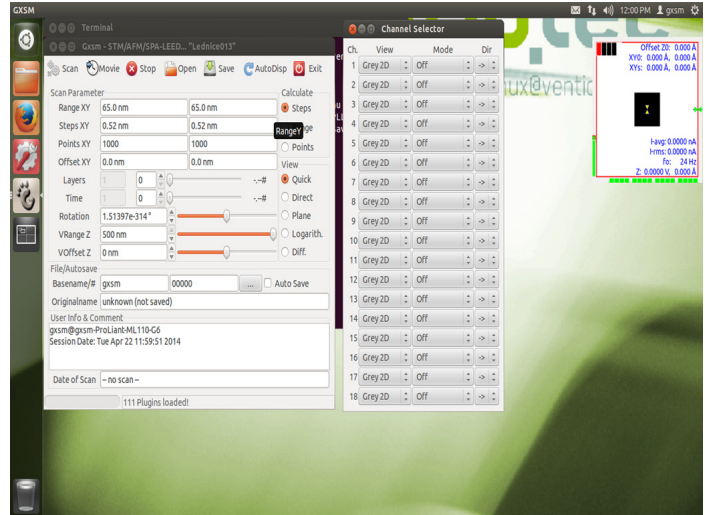

Figure 4 SPM connection preview

### 5. Model Setup of Proposed Multi Probe SPM

- In today's world, nanotechnology plays a vital role in the fields of science, engineering, agriculture, medicine & etc. Understanding interplay between various electrical, chemical and mechanical properties are very important for the development of new nano materials.
- Before production, it is essential to test and analysis of these nano materials. The available commercial systems are not versatile to carry out in situ real time measurements for multi domain nano scale properties.
- We identify lacunae to measure a multi domain property and its interplay analysis of nanomaterials. In our proposed technique, we develop a scanning probe microscopy system using MK-3 pro SPM controller with GXSM software to measure and analysis the multi domain properties of a nano material.
- The multi properties of nano materials can be analyzed without disturbing the sample position.
- Proposed system consist of two modes of analysis:
    - Multi-probe sequential analysis (electrical, mechanical, chemical etc.,)
    - Multi-probe simultaneous analysis (mechanical versus electrical, mechanical versus chemical, electrical versus chemical, electrical versus mechanical and etc.,)

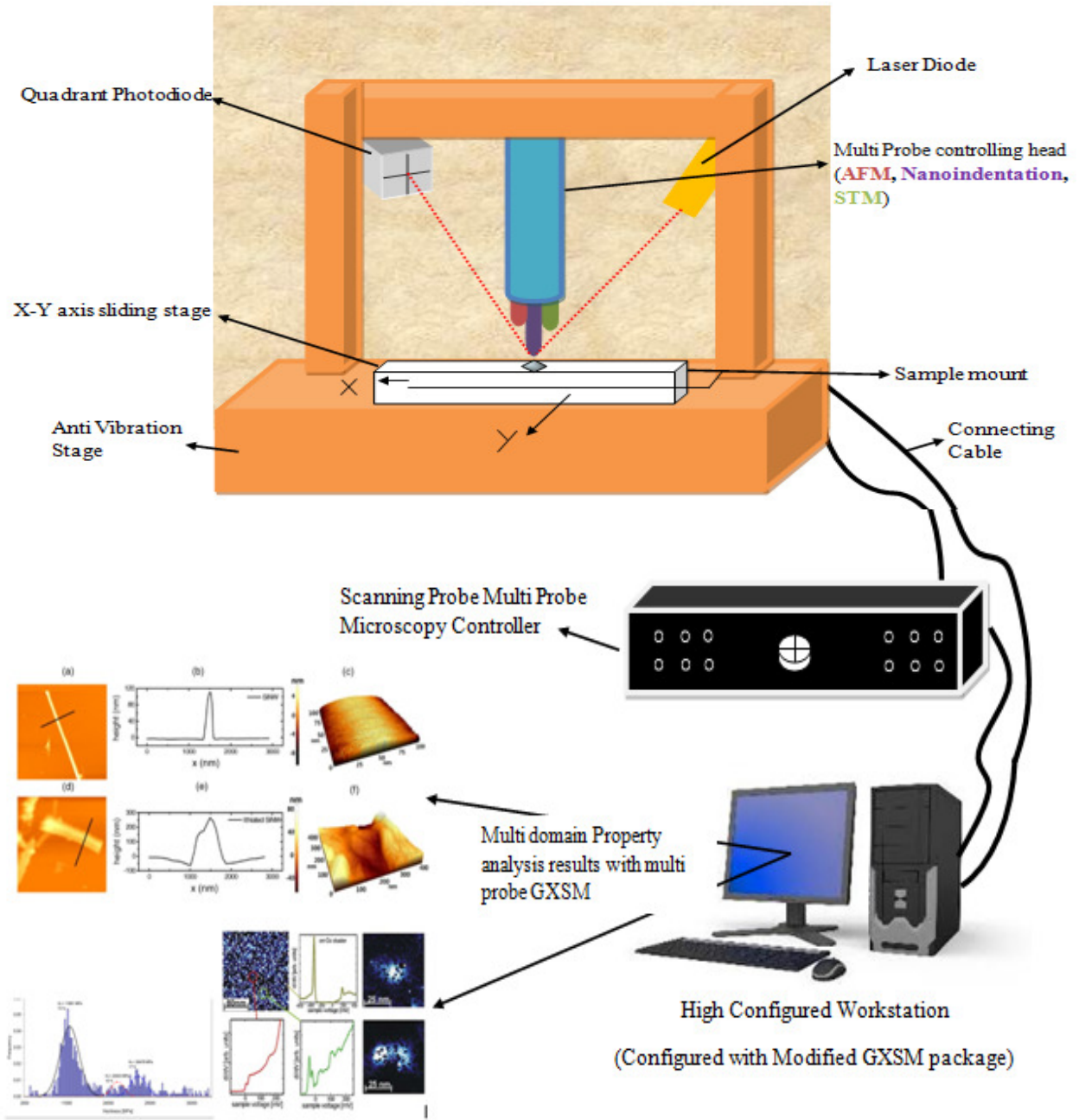

Figure 5 Multi-Probe model setup with open source SPM and GXSM package

## Conclusion

In this project the software module for connecting the GXSM user system to the DSP controller system has been designed and configured with the Linux based OS. The signal ranger controlling window has been verified with the installed Kernel.

The future work of this research work will be carried out by design, control and development of multi probe hardware setup as an advanced tool of Nanotechnology.

## Acknowledgement

We take immense pleasure to express my sincere and deep sense of gratitude to the DST Nano mission for their funding to this project. The authors thank all volunteers contributing to the Gxsm project and the Gxsm user community for many ideas and discussions. We also thank Mr. A. Alfred Raja Melvin for help with Linux installation.

## References

[1] P. Zahl, M. Bierkandt, S. Schr¨oder, and A. Klust, Rev.Sci. Inst. 74, 1222 (2003).
[2] P. Zahl, T. Wagner, R. Möller and A. Klust, "Open source scanning probe microscopy control software package Gxsm", J. Vac. Sci. Technol. B 28 (2010).
[3] P. Zahl and A. Klust, gxsm Software Project Homepage:URL: http://gxsm.sourceforge.net.
[4] P. Zahl, A. Klust, and S. Schr¨oder, The Gxsm Manual, Institut for Festk¨orperphysik, Hannover, Germany; Colorado School of Mines, Golden, CO, USA; University of Washington, Seattle, WA, USA (2002), Online available here: URL: http://gxsm.sourceforge.net, section Manual/Help.
[5] SoftdB, manufacturer of all the SignalRanger DSP Boards, URL: http://www.softdb.com.
[6] MK2-A810: Signal Ranger MK2 with Analog810 Module, URL: http://www.softdb.com/a-dsp_SPM_Controller.html.
[7] Gxsm Life CD/DVD Ubuntu System can be downloaded here: URL: http://www.ventiotec.com/linux/
[8] GXSM-Linux.iso, provided via Ventiotec Dolega Wagner GbR, Marler Straße 100-102, 45896 Gelsenkirchen, Germany.
[9] Unidata, NetCDF Homepage: URL: http://www.unidata.ucar.edu/packages/netcdf.
[10] Python scripting language: http://www.python.org.
[11] Piezodrive by Ing. B¨uro W. Reimann, Germany. Go to URL: http://gxsm.sourceforge.net and look at the section hardware for details.